\documentclass[conference]{IEEEtran}
\IEEEoverridecommandlockouts

\usepackage{cite}
\usepackage{amsmath,amssymb,amsfonts}
\usepackage{algorithmic}
\usepackage{graphicx}
\usepackage{textcomp}
\usepackage{xcolor}
\usepackage{booktabs}
\usepackage{tabularx}
\usepackage{array}
\usepackage{makecell}
\usepackage[hidelinks]{hyperref}
\def\BibTeX{{\rm B\kern-.05em{\sc i\kern-.025em b}\kern-.08em
    T\kern-.1667em\lower.7ex\hbox{E}\kern-.125emX}}
\begin{document}

\title{Trigger Value Selection in Clean-Label Malware Backdoors: An Effectiveness--Detectability Analysis}

\author{
\IEEEauthorblockN{Tri Budi Sabdo Utomo, Kazushi Okamoto, Koki Karube, Kei Harada, Atsushi Shibata}
\IEEEauthorblockA{Department of Informatics, Graduate School of Informatics and Engineering, The University of Electro-Communications}
\IEEEauthorblockA{1-5-1 Chofugaoka, Chofu, Tokyo 182-8585, Japan}
\IEEEauthorblockA{\{tribudi-utomo, kazushi, karubekoki, harada, shibata-atsushi\}@uec.ac.jp}
}

\IEEEaftertitletext{%
  \begin{minipage}{\textwidth}
    \footnotesize
    \raggedright
    \copyright\ 2026 IEEE.
    Personal use of this material is permitted.
    Permission from IEEE must be obtained for all other uses, in any current or future media, including reprinting/republishing this material for advertising or promotional purposes, creating new collective works, for resale or redistribution to servers or lists, or reuse of any copyrighted component of this work in other works.
    \par
  \end{minipage}
  \vspace{1em}
}

\maketitle

\begin{abstract}
Clean-label backdoor poisoning is a threat to machine learning-based malware classifiers because poisoned samples can be inserted into training data without changing their labels.
Previous explanation-guided attacks mainly focus on selecting important trigger features, but the values assigned to these features can also affect the attack behavior.
In this paper, we study trigger value selection as an independent design factor in clean-label backdoor attacks against feature-based malware classifiers.
We evaluate several value-selection strategies on EMBER 2018 and a Win64 PE subset of EMBER 2024 using LightGBM classifiers.
We compare rare-value, SHAP-based, frequency-bounded, co-occurrence-aware, and quantile-based selectors under random and distribution-based malware-similar benign sampling.
The results show that aggressive selectors such as MinPopulation and CountAbsSHAP can achieve high attack success rate, especially under distribution-based sampling, but they are also easier to detect by sanitization methods such as Isolation Forest.
In contrast, selectors such as CorrCountAbsSHAP and Quantile95 can reduce poison recall while keeping useful attack effectiveness.
These findings show that trigger value selection is not merely an implementation detail, but an important design choice that shapes the effectiveness--detectability profile of clean-label malware backdoor attacks.
\end{abstract}


\section{Introduction}
Machine learning has substantially improved malware detection by enabling classifiers to distinguish unseen software samples as malicious or benign from static or behavioral features.
Datasets such as EMBER provide feature representations of Windows Portable Executable (PE) files and continue to support research on static malware classification~\cite{ember2018}.
However, malware evolves over time, causing classifier performance to degrade and requiring detection systems to be updated or trained with newly collected samples~\cite{concept_drift}.
This retraining process creates a new attack surface, especially when the training pipeline relies on external or crowdsourced threat feeds.
Such feeds can become injection points for clean-label backdoor poisoning, in which attackers insert poisoned samples into training data without controlling the labeling process~\cite{severi}.
In this attack, the attackers modify benign-labeled training samples by embedding trigger patterns, such as sets of feature-value assignments, while preserving their labels~\cite{severi,gu_badnets,pmsb}.
The classifier may then learn to associate the trigger pattern with the benign class, and at inference time, malware samples containing the same trigger may be predicted as benign~\cite{severi,pmsb}.
This attack is difficult to notice when the backdoored model still performs normally on clean samples.

Severi et al. introduced an explanation-guided clean-label backdoor poisoning attack against malware classifiers, showing that feature-based malware detectors can be manipulated through carefully designed trigger patterns~\cite{severi}.
Their attack uses explanation scores, such as Shapley additive explanations (SHAP) values, to select influential trigger features and then assigns values to those features so that poisoned benign samples teach the classifier to treat the trigger as benign.
They also proposed several value-selection strategies, including MinPopulation, CountAbsSHAP, and the greedy CombinedSHAP selector~\cite{severi}.
These strategies show that value selection is a substantive design choice: rare or low-density values may create stronger backdoor signals, whereas benign-like or co-occurring values may reduce detectability at the cost of attack effectiveness.

Previous studies have demonstrated the effectiveness of clean-label backdoor poisoning attacks against malware classifiers. 
However, the role of trigger value selection remains less systematically analyzed. 
Existing work has examined several factors, including poisoning rate, trigger feature selection, sample selection, and data sanitization defenses~\cite{severi,ho_sanitization}.
However, once the trigger features are fixed, the values assigned to those features can still strongly influence both attack success and detectability.
For example, rare values may make the trigger easier for the model to learn, but they can also create distinctive patterns that are easier to remove by sanitization methods.
In contrast, benign-like or co-occurring values may blend better into the benign distribution but may produce a weaker backdoor effect.
These observations suggest that trigger value selection changes the effectiveness--detectability relationship and should be analyzed as an independent design factor.

In this work, we investigate trigger value selection as an independent design factor in clean-label backdoor poisoning attacks against malware classifiers.
After fixing the trigger features, we compare rare-value, SHAP-based, frequency-bounded, co-occurrence-aware, prototype-based, and quantile-based value-selection strategies.
We evaluate these strategies in terms of clean accuracy, attack success, and sanitizer-based detectability.
We also examine how malware-similar benign sampling, such as distribution-based distance sampling following Poisoning Malware-Similar Benignware (PMSB)~\cite{pmsb}, affects the effectiveness--detectability relationship.
Specifically, we address the following research questions.
\begin{enumerate}
    \item[RQ1)] How do trigger value-selection strategies affect the effectiveness of the backdoor poisoning attack?
    \item[RQ2)] How do trigger value-selection strategies affect detectability under sanitization defenses?
    \item[RQ3)] How does malware-similar benign sampling affect the effectiveness--detectability relationship?
\end{enumerate}

The contributions of this work are as follows.
\begin{enumerate}
    \item We isolate trigger value selection as a design factor distinct from trigger feature selection and poisoning-sample selection.
    \item We compare rare-value, SHAP-based, frequency-bounded, co-occurrence-aware, prototype-based, and quantile-based value selectors.
    \item We evaluate how these selectors change both attack effectiveness and sanitizer-based detectability, including under random and malware-similar benign sampling.
\end{enumerate}

The remainder of this paper is organized as follows.
Section II reviews background and related work.
Section III describes the attack pipeline and value-selection strategies.
Section IV presents the experimental setup.
Section V reports the results and discusses the effectiveness--detectability relationship.
\section{Background and Related Work}
This section reviews clean-label backdoor poisoning attacks, sampling strategies, and defenses that are relevant to our analysis of trigger value selection.

\subsection{Clean-Label Backdoor Poisoning Attacks}
Severi et al. introduced an explanation-guided clean-label backdoor poisoning attack against malware classifiers~\cite{severi}.
Their method uses SHAP values to select influential trigger features and assigns trigger values using strategies such as MinPopulation, CountAbsSHAP, and CombinedSHAP.
The poisoned samples retain their benign labels, causing the trained model to associate the trigger with the benign class while maintaining normal performance on clean samples.
Other studies have examined clean-label malware backdoors under different assumptions, including black-box attacks using PE surface information and selective backdoors with realizable triggers~\cite{zheng_clean_label,jigsaw}.

\subsection{Sampling Strategies in Clean-Label Malware Poisoning}
In addition to selecting trigger features and trigger values, choosing which samples to poison can also affect the effectiveness of malware backdoor poisoning attacks.
Previous work has argued that benignware samples are not all equally useful for poisoning. Some benign samples are more similar to malware samples than others.
Selecting these malware-similar benign samples as poisoning targets can improve the effectiveness of the backdoor attack.
This approach uses similarity metrics such as feature-based distance, distribution-based distance, and contribution-space distance~\cite{pmsb}.
We use sampling strategies such as PMSB to evaluate whether the observed effectiveness--detectability relationship remains consistent under different poisoning-target selection strategies.

\subsection{Defenses and Detectability}
Previous studies have proposed defenses against clean-label backdoor poisoning attacks on malware classifiers.
Severi et al. evaluated sanitization methods such as Isolation Forest, Spectral Signature, and clustering-based detection~\cite{severi}.
Other studies have investigated data sanitization defenses based on ensemble model disagreement, nested training, dimensionality reduction, clustering, and model-agnostic mitigation strategies~\cite{ho_sanitization,venkatesan_nids,countermeasures,model_agnostic_mitigation,reddy_improved_nt,partition,yudin_retraining}.
These defenses suggest that poisoned samples can be detected through multiple signals, including data anomalies, shared directions in feature space, prediction disagreement, and clustering behavior.
In this work, we use representative sanitization defenses to measure how different value-selection strategies affect the detectability of poisoned samples.
\section{Methodology}
\subsection{Attack Pipeline}
We study clean-label poisoning attacks against feature-based malware classifiers.
Let $\boldsymbol{x} \in \mathcal{X}$ denote a feature vector extracted from a benign or malware sample, and let $y \in \{0, 1\}$ denote its label, where 0 and 1 indicate benign and malware, respectively.
A trigger is represented as a set of feature-value assignments $T = \{(f_{1}, v_{1}), (f_{2}, v_{2}), \cdots, (f_{k}, v_{k})\}$, where $f_{i}$ is a selected trigger feature, and $v_{i}$ is the value assigned to that feature.
For an input sample $\boldsymbol{x}$, trigger injection overwrites the original values of the selected features with the corresponding trigger values, producing a triggered sample $\boldsymbol{x}_{T}$.

\begin{figure}[t]
    \centering
    \includegraphics[width=\hsize]{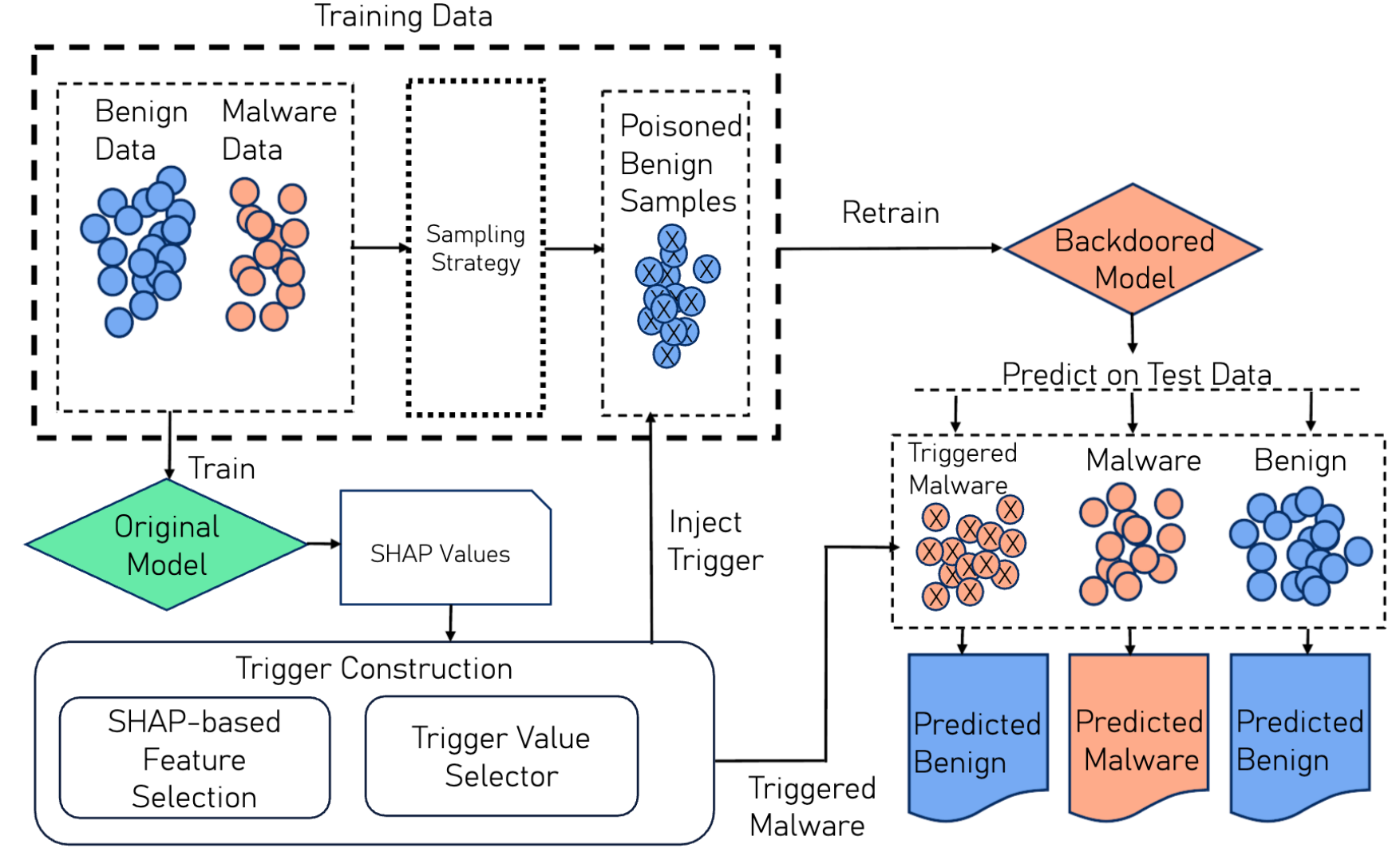}
    \caption{Overview of the clean-label backdoor poisoning pipeline.}
    \label{fig:attack_pipeline}
\end{figure}

Fig.~\ref{fig:attack_pipeline} shows the overall attack pipeline used in this work.
Our pipeline follows the explanation-guided clean-label backdoor attack setting~\cite{severi}.
We first train a clean baseline classifier on the original training data and then compute feature importance scores using SHAP values.
A fixed set of trigger features is selected based on the largest aggregated absolute SHAP values across training samples, subject to feature-space feasibility constraints so that only feasible features are used to construct the backdoor trigger.
For each experiment, we use one value-selection strategy to choose the trigger values from the original training data for the selected trigger features. 
The resulting trigger is injected into a subset of benign training samples without changing their labels.
The classifier is then retrained on the poisoned training set to obtain the backdoored model.
Finally, we evaluate the backdoored model on clean test samples and triggered malware samples.
The attack objective is to cause triggered malware samples to evade detection while preserving normal performance on clean samples; accordingly, we use attack success rate as the primary attack-effectiveness metric.

\subsection{Trigger Value Selection Strategies}
The main focus of this work is trigger value selection.
After the set of trigger features $F_{T} = \{f_{1}, f_{2}, \cdots, f_{k}\}$ is fixed, a value selector determines which value $v_{i}$ should be assigned to each selected feature $f_{i}$.
In addition to the Severi-derived selectors~\cite{severi}, we introduce five extended selectors: FreqBoundedSigned, LowSignedSHAP, CorrCountAbsSHAP, BenignPrototype, and QuantileDistribution.
These selectors are designed to test whether signed SHAP direction, frequency bounds, co-occurrence support, benign prototypes, or empirical quantiles can change the effectiveness--detectability relationship.
We group the value selectors according to the main signal used by each selector: frequency or distribution, SHAP contribution, and co-occurrence or prototype support.
These categories are not mutually exclusive; some selectors combine multiple signals, such as frequency and SHAP magnitude, or signed SHAP contribution and co-occurrence.
Table~\ref{tab:value_selectors} summarizes the value-selection strategies evaluated in this work.

\subsubsection{Frequency- and Quantile-based Selectors}
The selectors choose values according to how common, rare, or distributionally positioned they are in benign data.
Rare-value selection places the trigger in a sparse region of the feature space, following the intuition of the MinPopulation method~\cite{severi}.
Such values may give the trigger stronger leverage over the decision boundary, but they may also create more distinctive patterns that are easier to detect.
Frequency-bounded and quantile-based selectors are included to test whether controlling the rarity or empirical position of trigger values changes the effectiveness--detectability profiles.
For QuantileDistribution, we evaluated Quantile05, Quantile50, and Quantile95, and report Quantile95 in the main tables because it gave the strongest and most consistent results.

\subsubsection{SHAP-based Selectors}
The selectors use feature attribution scores to estimate how candidate values contribute to the classifier's decision.
Following the intuition of the CountAbsSHAP method~\cite{severi}, magnitude-based selectors use feature importance scores to identify values with low absolute class-specific contributions.
For malware-class explanations, negative SHAP values push the classifier's decision toward the benign class.
Signed SHAP selectors use this direction of contribution and select values that are expected to push predictions toward the benign class.
We evaluate whether using signed SHAP values produces stronger or less detectable triggers than magnitude-only criteria.

\subsubsection{Co-occurrence- and Prototype-based Selectors}
The selectors consider whether selected feature-value pairs naturally appear together in benign samples.
This follows the intuition of the CombinedSHAP method~\cite{severi}, which greedily selects feature-value pairs that remain consistent with existing benign samples.
Unlike independent value selection, co-occurrence-aware selection aims to preserve relationships among selected feature values so that the trigger better blends into the benign distribution.
Previous work has noted that independently selected values may not correspond to naturally occurring benign patterns, while CombinedSHAP is designed to align with existing benign-oriented regions~\cite{severi,ho_sanitization}.

\begin{table*}[!t]
    \centering
    \caption{Summary of trigger value-selection strategies.}
    \label{tab:value_selectors}
    \begin{tabularx}{\textwidth}{>{\raggedright\arraybackslash}p{2.2cm}
    >{\raggedright\arraybackslash}p{3.2cm}
    >{\raggedright\arraybackslash}X
    >{\raggedright\arraybackslash}p{3.2cm}}
        \toprule
        \textbf{Source} & \textbf{Selector} & \textbf{Main principle} & \textbf{Category} \\
        \midrule
        Severi-derived & MinPopulation & Selects rare observed values. & Frequency / rarity \\
        Severi-derived & CountAbsSHAP & Balances value frequency and low absolute SHAP contribution. & Frequency + SHAP magnitude \\
        Severi-derived & CombinedSHAP & Greedily selects benign-oriented co-occurring feature-value pairs. & Signed SHAP + co-occurrence \\
        This work & FreqBoundedSigned & Selects benign-oriented values within a frequency range. & Frequency + signed SHAP \\
        This work & LowSignedSHAP & Selects values with a benign-oriented signed contribution. & Signed SHAP \\
        This work & CorrCountAbsSHAP & Considers low absolute SHAP values with co-occurrence support. & SHAP magnitude + co-occurrence \\
        This work & BenignPrototype & Selects values from representative benign patterns. & Prototype / co-occurrence \\
        This work & QuantileDistribution & Selects values from fixed empirical distribution positions. & Distribution baseline \\
        \bottomrule
    \end{tabularx}
\end{table*}

\section{Experimental Setup}
We instantiate the explanation-guided clean-label attack pipeline described in Section III using LightGBM classifiers~\cite{lightgbm}.
Within each experimental condition, the classifier family, trigger-feature selection procedure, and poisoning protocol are fixed, while only the trigger value-selection strategy is varied.
This design isolates the effect of value selection on attack effectiveness and sanitizer-based detectability.

We use EMBER 2018~\cite{ember2018} and the Win64 PE subset of EMBER 2024~\cite{ember2024}.
The EMBER 2018 source contains 800,000 training rows, of which 600,000 are labeled and 200,000 are unlabeled.
We sample 20\% of the labeled training pool using balanced stratified sampling, resulting in 120,000 training rows (60,000 benign and 60,000 malware); the unlabeled rows are not used.
The EMBER 2018 test split contains 200,000 labeled rows (100,000 benign and 100,000 malware), all of which are used for evaluation.
For EMBER 2024 Win64, the 20\% stratified training subset contains 208,000 labeled rows (104,000 benign and 104,000 malware), and the complete test split contains 240,000 labeled rows (120,000 benign and 120,000 malware).
The EMBER 2018 and EMBER 2024 Win64 representations contain 2,381 and 2,568 features, respectively.

We focus on Win64 PE files because our attack setting targets static Windows PE malware classifiers.
Following prior explanation-guided malware backdoor work, we restrict trigger construction to problem-space-conservative features, retaining 17 trigger features for EMBER 2018 and 19 for EMBER 2024 Win64.
Because this study is conducted in feature space, we do not validate the selected triggers by modifying executable PE binaries.
For poisoning-sample selection, we compare random benign sampling with distribution-based distance sampling following PMSB~\cite{pmsb}.
The main effectiveness--detectability comparison uses a 1\% poison rate.
The implementation also supports higher poison rates, but those settings are outside the scope of the analysis reported in this paper.
The evaluated value selectors are summarized in Table~\ref{tab:value_selectors}.

We measure attack effectiveness using clean accuracy and attack success rate (ASR), where ASR is the fraction of triggered malware samples that are detected by the clean model but classified as benign by the backdoored model.
For sanitizer-based detectability, we use poison recall as the primary metric; the numbers of poisoned and benign rows removed are retained in the experiment artifacts for supplementary analysis.
A high poison recall is desirable for the defender because more poisoned samples are removed, but it is undesirable for the attacker because it can prevent the backdoor from being learned during retraining.

We evaluate Isolation Forest~\cite{isolation_forest}, HDBSCAN~\cite{hdbscan}, and Spectral Signature~\cite{spectral_signature} on standardized hybrid features using the top 32 selected features.
The settings are contamination $0.02$ for Isolation Forest, 2\% removal for Spectral Signature, and HDBSCAN parameters of 0.5\% minimum cluster size, 0.1\% minimum samples, 10\% maximum cluster size, and 0.2 minimum keep probability.
Due to space limitations, the main tables report Isolation Forest as a representative sanitizer baseline, while HDBSCAN and Spectral Signature are used in the broader experiment to check whether the qualitative trends are consistent across sanitizers.
\section{Results}
\subsection{Effect of Value Selection on ASR}
Tables~\ref{tab:ember2018_1p_if} and \ref{tab:ember2024_1p_if} report the mean results at a 1\% poison rate using Isolation Forest as a compact detectability baseline.
Each reported experimental condition was repeated three times, and the tables report the corresponding mean and standard deviation.
The main text focuses on the 1\% Isolation Forest setting because it compactly shows how value selection changes ASR and poison recall.

The results indicate that trigger value selection strongly affects attack success even when the trigger feature set is fixed.
The Severi-derived selectors, especially MinPopulation and CountAbsSHAP, achieve very high ASR under distribution-based distance sampling on both datasets.
On EMBER 2018, they achieve 97.94\% and 98.87\% ASR, respectively, while on EMBER 2024 Win64 they achieve 99.14\% and 99.04\%.
This confirms that rare-value and low-SHAP-contribution value selection can produce effective triggers.

Among the extended selectors, CorrCountAbsSHAP also maintains useful ASR under distribution-based sampling, reaching 79.33\% on EMBER 2018 and 98.85\% on EMBER 2024 Win64.
Quantile95 achieves 93.74\% ASR on EMBER 2018 and 92.73\% on EMBER 2024 Win64. By contrast, CombinedSHAP and BenignPrototype generally achieve lower ASR, particularly under random sampling.
Overall, the results indicate that the selected trigger values substantially change attack effectiveness, even when the trigger features and poisoning protocol are held constant.

\begin{table}[!t]
    \centering
    \caption{EMBER 2018 results at a 1\% poison rate using Isolation Forest. Values are mean $\pm$ standard deviation across 3 runs.}
    \label{tab:ember2018_1p_if}
    \setlength{\tabcolsep}{3.0pt}
    \renewcommand{\arraystretch}{1.08}
    \begin{tabular}{lrrr}
        \toprule
        \textbf{Selector} &
        \makecell{\textbf{ASR}\\($\%$)} &
        \makecell{\textbf{Clean accuracy}\\($\%$)} &
        \makecell{\textbf{Poison recall}\\($\%$)} \\
        \midrule
        \multicolumn{4}{l}{\textit{Random sampling}} \\
        MinPopulation*      & 77.71 $\pm$ 14.52 & 97.80 $\pm$ 0.01 & 100.00 $\pm$ 0.00 \\
        CountAbsSHAP*       & 69.68 $\pm$ 6.34  & 97.70 $\pm$ 0.10 & 100.00 $\pm$ 0.00 \\
        CombinedSHAP*       & 46.87 $\pm$ 8.02  & 97.67 $\pm$ 0.08 & 0.08 $\pm$ 0.12 \\
        FreqBoundedSigned   & 60.74 $\pm$ 8.38  & 97.59 $\pm$ 0.18 & 89.00 $\pm$ 2.83 \\
        LowSignedSHAP       & 74.75 $\pm$ 12.59 & 97.69 $\pm$ 0.11 & 98.50 $\pm$ 0.94 \\
        CorrCountAbsSHAP    & 28.75 $\pm$ 10.83 & 97.69 $\pm$ 0.07 & 3.63 $\pm$ 2.30 \\
        BenignPrototype     & 5.81 $\pm$ 1.64  & 97.66 $\pm$ 0.04 & 0.04 $\pm$ 0.06 \\
        Quantile95          & 45.55 $\pm$ 8.81  & 97.61 $\pm$ 0.07 & 26.33 $\pm$ 6.72 \\
        \midrule
        \multicolumn{4}{l}{\textit{Distribution-based sampling}} \\
        MinPopulation*      & 97.94 $\pm$ 0.58 & 97.79 $\pm$ 0.09 & 100.00 $\pm$ 0.00 \\
        CountAbsSHAP*       & 98.87 $\pm$ 0.21 & 97.92 $\pm$ 0.08 & 99.71 $\pm$ 0.41 \\
        CombinedSHAP*       & 86.86 $\pm$ 1.07 & 97.90 $\pm$ 0.14 & 0.00 $\pm$ 0.00 \\
        FreqBoundedSigned   & 95.35 $\pm$ 2.51 & 97.99 $\pm$ 0.02 & 90.33 $\pm$ 6.72 \\
        LowSignedSHAP       & 88.74 $\pm$ 0.06 & 97.82 $\pm$ 0.19 & 98.63 $\pm$ 1.94 \\
        CorrCountAbsSHAP    & 79.33 $\pm$ 18.94 & 97.70 $\pm$ 0.16 & 1.13 $\pm$ 1.12 \\
        BenignPrototype     & 42.66 $\pm$ 17.93 & 97.67 $\pm$ 0.26 & 0.00 $\pm$ 0.00 \\
        Quantile95          & 93.74 $\pm$ 1.76 & 97.78 $\pm$ 0.11 & 6.33 $\pm$ 5.07 \\
        \bottomrule
    \end{tabular}
    \vspace{0.5mm}
    \raggedright
    * indicates selectors derived from Severi et al.
\end{table}

\begin{table}[!t]
    \centering
    \caption{EMBER 2024 Win64 results at a 1\% poison rate using Isolation Forest. Values are mean $\pm$ standard deviation across 3 runs.}
    \label{tab:ember2024_1p_if}
    \setlength{\tabcolsep}{3.0pt}
    \renewcommand{\arraystretch}{1.08}
    \begin{tabular}{lrrr}
        \toprule
        \textbf{Selector} &
        \makecell{\textbf{ASR}\\($\%$)} &
        \makecell{\textbf{Clean accuracy}\\($\%$)} &
        \makecell{\textbf{Poison recall}\\($\%$)} \\
        \midrule
        \multicolumn{4}{l}{\textit{Random sampling}} \\
        MinPopulation*      & 15.74 $\pm$ 1.96 & 98.58 $\pm$ 0.06 & 100.00 $\pm$ 0.00 \\
        CountAbsSHAP*       & 15.42 $\pm$ 0.31 & 98.58 $\pm$ 0.07 & 100.00 $\pm$ 0.00 \\
        CombinedSHAP*       & 7.24 $\pm$ 0.57  & 98.53 $\pm$ 0.04 & 1.23 $\pm$ 0.71 \\
        FreqBoundedSigned   & 15.49 $\pm$ 0.32 & 98.57 $\pm$ 0.09 & 83.56 $\pm$ 6.12 \\
        LowSignedSHAP       & 17.43 $\pm$ 3.26 & 98.58 $\pm$ 0.05 & 98.56 $\pm$ 1.09 \\
        CorrCountAbsSHAP    & 11.38 $\pm$ 0.78 & 98.55 $\pm$ 0.06 & 2.81 $\pm$ 2.14 \\
        BenignPrototype     & 2.67 $\pm$ 0.06  & 98.57 $\pm$ 0.01 & 0.00 $\pm$ 0.00 \\
        Quantile95          & 9.71 $\pm$ 1.77  & 98.52 $\pm$ 0.17 & 22.79 $\pm$ 8.97 \\
        \midrule
        \multicolumn{4}{l}{\textit{Distribution-based sampling}} \\
        MinPopulation*      & 99.14 $\pm$ 0.06 & 98.83 $\pm$ 0.03 & 100.00 $\pm$ 0.00 \\
        CountAbsSHAP*       & 99.04 $\pm$ 0.00 & 98.83 $\pm$ 0.04 & 100.00 $\pm$ 0.00 \\
        CombinedSHAP*       & 72.34 $\pm$ 17.51 & 98.77 $\pm$ 0.04 & 2.57 $\pm$ 0.44 \\
        FreqBoundedSigned   & 95.79 $\pm$ 0.02 & 98.83 $\pm$ 0.00 & 78.87 $\pm$ 18.53 \\
        LowSignedSHAP       & 93.39 $\pm$ 0.15 & 98.85 $\pm$ 0.01 & 99.01 $\pm$ 1.26 \\
        CorrCountAbsSHAP    & 98.85 $\pm$ 0.59 & 98.82 $\pm$ 0.04 & 10.34 $\pm$ 3.13 \\
        BenignPrototype     & 67.19 $\pm$ 19.93 & 98.78 $\pm$ 0.04 & 0.00 $\pm$ 0.00 \\
        Quantile95          & 92.73 $\pm$ 5.78 & 98.71 $\pm$ 0.15 & 29.09 $\pm$ 1.77 \\
        \bottomrule
    \end{tabular}
    \vspace{0.5mm}
    \raggedright
    * indicates selectors derived from Severi et al.
\end{table}

\subsection{Detectability Profiles}
\begin{figure}[!t]
    \centering
    \includegraphics[width=\columnwidth]{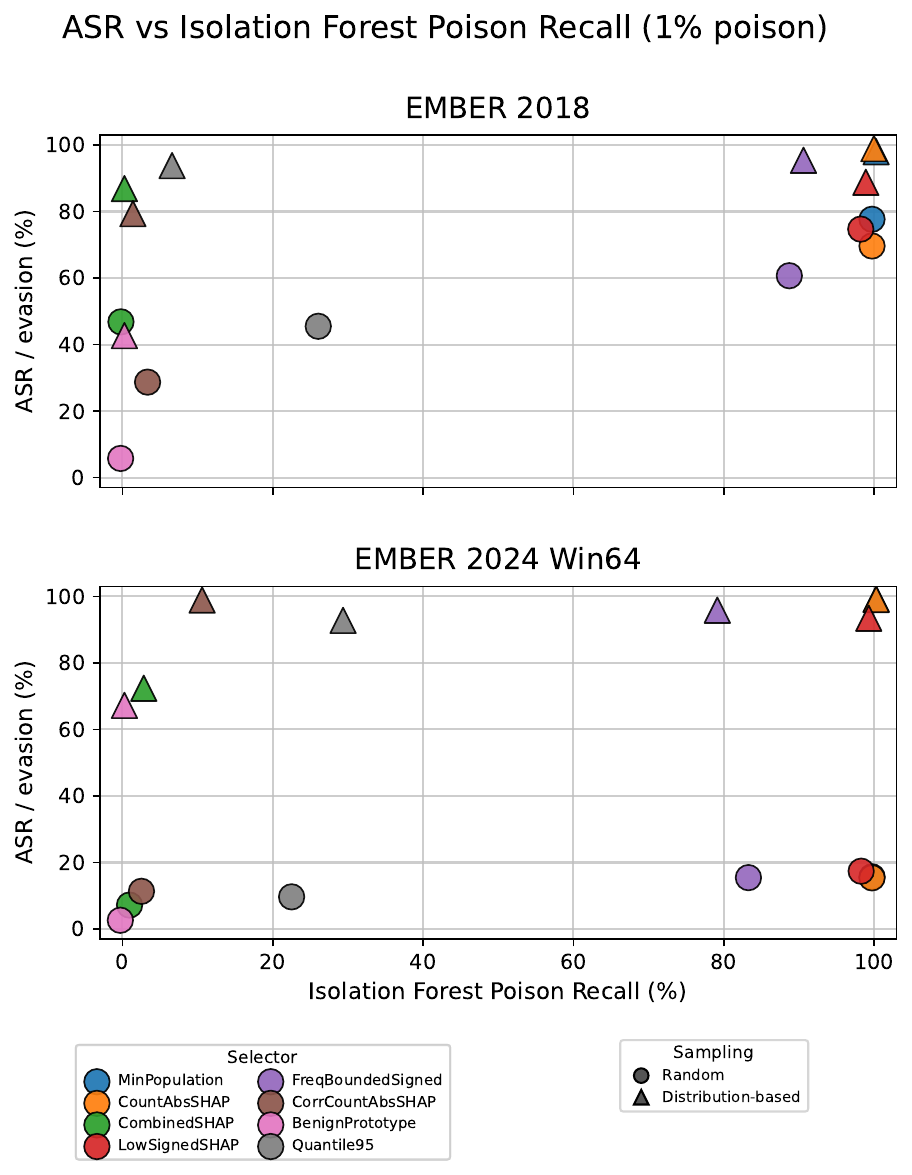}
    \caption{Attack effectiveness versus Isolation Forest detectability at 1\% poison rate. ASR/evasion is plotted against poison recall for EMBER 2018 and EMBER 2024 Win64. Each point represents the mean across 3 repeated runs.}
    \label{fig:asr-defense-profile}
\end{figure}

Fig.~\ref{fig:asr-defense-profile} visualizes the effectiveness--detectability profile by plotting ASR against Isolation Forest poison recall.
Points in the upper-right region are effective but easy to detect, whereas points in the upper-left region are effective and harder to detect.

As shown in Tables~\ref{tab:ember2018_1p_if} and~\ref{tab:ember2024_1p_if}, MinPopulation and CountAbsSHAP occupy the high-ASR and high-recall region under distribution-based distance sampling.
Under random sampling, they remain highly exposed to Isolation Forest, but their ASR can be much lower, especially on EMBER 2024 Win64.
In contrast, CombinedSHAP, CorrCountAbsSHAP, BenignPrototype, and Quantile95 achieve lower poison recall but differ in how much ASR they preserve.
CombinedSHAP and BenignPrototype can reduce poison recall, but this reduction often comes with substantial ASR loss.
CorrCountAbsSHAP is more attractive because it preserves high ASR under distribution-based distance sampling while keeping poison recall much lower than the strongest Severi-derived baselines.
FreqBoundedSigned shows an intermediate profile, retaining relatively high ASR under distribution-based sampling while remaining moderately exposed to sanitization.
LowSignedSHAP is closer to the aggressive selectors, achieving high ASR but also very high poison recall on both datasets.

Compared with CountAbsSHAP, CorrCountAbsSHAP reduces poison recall from 99.71\% to 1.13\% on EMBER 2018 and from 100.00\% to 10.34\% on EMBER 2024 Win64 under distribution-based sampling, while retaining ASR of 79.33\% and 98.85\%, respectively.
Quantile95 shows a favorable effectiveness–detectability profile.
It achieves 93.74\% ASR with 6.33\% poison recall on EMBER 2018 and 92.73\% ASR with 29.09\% poison recall on EMBER 2024 Win64. In comparison, MinPopulation and CountAbsSHAP achieve approximately 98–99\% ASR but approximately 99–100\% poison recall.

We also evaluated HDBSCAN and Spectral Signature to check whether the observations are specific to Isolation Forest.
Although poison-recall levels vary across sanitizers and datasets, HDBSCAN and Spectral Signature generally reproduce the same trend: MinPopulation, CountAbsSHAP, and LowSignedSHAP are more exposed, whereas CorrCountAbsSHAP and Quantile95 retain useful ASR with lower poison recall.
For example, under distribution-based sampling, HDBSCAN poison recall ranges from 64.13--76.75\% for MinPopulation and CountAbsSHAP, but only 26.50--43.70\% for CorrCountAbsSHAP and Quantile95 across the two datasets.
Spectral Signature shows an even stronger contrast, with 58.71--97.67\% recall for the former selectors versus 0.79--9.81\% for the latter.
Among the extended selectors, CorrCountAbsSHAP and Quantile95 therefore provide the clearest effectiveness--detectability trade-off, preserving attack effectiveness while reducing the fraction of poisoned samples identified by sanitization.
Evaluating ASR alone can therefore be misleading, because a highly effective attack may still be neutralized if most poisoned samples are removed before retraining.

\subsection{Effect of Sampling Strategy}
Sampling strategy also has a strong effect on attack effectiveness.
Distribution-based distance sampling consistently improves ASR compared with random sampling, especially on EMBER 2024 Win64.
This is consistent with PMSB-style sampling, where poisoning malware-similar benign samples makes the trigger easier for the model to associate with the benign class because these samples are closer to the decision boundary.

However, sampling strategy does not remove the role of value selection.
Even under distribution-based distance sampling, selectors still differ substantially in both ASR and poison recall.
This means that sample selection and value selection control different parts of the attack.
PMSB-style sampling determines which benign samples are poisoned, whereas value selection determines which trigger pattern is learned from those poisoned samples.
Thus, PMSB-style sampling can strengthen the attack, but the resulting effectiveness–detectability profile still depends on the selected trigger values.
\section{Conclusion}
This paper studied trigger value selection as an independent design factor in clean-label backdoor poisoning attacks against feature-based malware classifiers.
Regarding attack effectiveness (RQ1), our results show that the values assigned to fixed trigger features strongly affect ASR.
Across EMBER 2018 and EMBER 2024 Win64, aggressive selectors such as MinPopulation and CountAbsSHAP achieved high ASR, particularly under distribution-based distance sampling.
CorrCountAbsSHAP and Quantile95 also maintained useful attack effectiveness, showing that value selection can change the strength of the backdoor even when the trigger feature set is fixed.

Regarding sanitizer-based detectability (RQ2), our results show that trigger value selection also affects poison recall.
MinPopulation and CountAbsSHAP achieved high ASR, but they were also detected at high rates by Isolation Forest. By contrast, CorrCountAbsSHAP and Quantile95 reduced poison recall while preserving useful attack effectiveness.
These results show that trigger value selection changes the effectiveness--detectability relationship.
The strongest trigger is not necessarily the most practical if poisoned samples can be removed by sanitization.

Regarding malware-similar benign sampling (RQ3), distribution-based distance sampling generally improved ASR, especially on EMBER 2024 Win64, but it did not remove the effect of value selection. 
Sample selection and value selection are therefore complementary: one determines which benign samples are poisoned, while the other determines the trigger pattern learned by the classifier. 
Overall, trigger value selection is a core design choice that shapes the effectiveness--detectability profile of clean-label malware backdoors.

This study has several limitations. It focuses on feature-space trigger value selection, does not realize triggers in executable PE binaries, and evaluates the proposed selectors only on LightGBM-based classifiers and two PE malware datasets.
Future work should study problem-space realizations, broader models and datasets, stronger end-to-end defenses, and value selectors that jointly optimize attack effectiveness and detectability.


\begin{thebibliography}{00}

\bibitem{ember2018}
H. S. Anderson and P. Roth,
``EMBER: An Open Dataset for Training Static PE Malware Machine Learning Models,''
arXiv preprint arXiv:1804.04637, 2018,
doi: 10.48550/arXiv.1804.04637.

\bibitem{concept_drift}
R. Jordaney, K. Sharad, S. K. Dash, Z. Wang, D. Papini, I. Nouretdinov, and L. Cavallaro,
``Transcend: Detecting Concept Drift in Malware Classification Models,''
in \textit{Proc. 26th USENIX Security Symposium}, pp.625--642, 2017.

\bibitem{ember2024}
R. J. Joyce, G. Miller, P. Roth, R. Zak, E. Z.-Williams, H. Anderson, E. Raff, and J. Holt,
``EMBER2024: A Benchmark Dataset for Holistic Evaluation of Malware Classifiers,''
in \textit{Proc. 31st ACM SIGKDD Conference on Knowledge Discovery and Data Mining (KDD)}, pp.5516--5526, 2025,
doi: 10.1145/3711896.3737431.

\bibitem{lightgbm}
G. Ke, Q. Meng, T. Finley, T. Wang, W. Chen, W. Ma, Q. Ye, and T.-Y. Liu,
``LightGBM: A Highly Efficient Gradient Boosting Decision Tree,''
in \textit{Proc. 31st International Conference on Neural Information Processing Systems (NeurIPS)}, pp.3149--3157, 2017.

\bibitem{shap}
S. M. Lundberg and S.-I. Lee,
``A Unified Approach to Interpreting Model Predictions,''
in \textit{Proc. 31st International Conference on Neural Information Processing Systems (NeurIPS)}, pp.4768--4777, 2017.

\bibitem{severi}
G. Severi, J. Meyer, S. Coull, and A. Oprea,
``Explanation-Guided Backdoor Poisoning Attacks Against Malware Classifiers,''
in \textit{Proc. 30th USENIX Security Symposium}, pp.1487--1504, 2021.

\bibitem{gu_badnets}
T. Gu, B. D.-Gavitt, and S. Garg,
``BadNets: Identifying Vulnerabilities in the Machine Learning Model Supply Chain,''
arXiv preprint arXiv:1708.06733, 2017,
doi: 10.48550/arXiv.1708.06733.

\bibitem{pmsb}
X. Wang, Y. Feng, B. Bi, Y. Cao, Z. Jin, X. Liu, Y. Liu, and Y. Li,
``Not All Benignware Are Alike: Enhancing Clean-Label Attacks on Malware Classifiers,''
in \textit{Proc. ACM Web Conference 2025 (WWW)}, pp.2053--2063, 2025,
doi: 10.1145/3696410.3714552.

\bibitem{ho_sanitization}
S. Ho, A. Reddy, S. Venkatesan, R. Izmailov, R. Chadha, and A. Oprea,
``Data Sanitization Approach to Mitigate Clean-Label Attacks Against Malware Detection Systems,''
in \textit{Proc. 2022 IEEE Military Communications Conference (MILCOM)}, pp.993--998, 2022,
doi: 10.1109/MILCOM55135.2022.10017768.

\bibitem{reddy_improved_nt}
A. Reddy, S. Venkatesan, R. Izmailov, and A. Oprea,
``An Improved Nested Training Approach to Mitigate Clean-label Attacks against Malware Classifiers,''
in \textit{Proc. 2023 IEEE Military Communications Conference (MILCOM)}, pp.703--709, 2023,
doi: 10.1109/MILCOM58377.2023.10356253.

\bibitem{venkatesan_nids}
S. Venkatesan, H. Sikka, R. Izmailov, R. Chadha, A. Oprea, and M. J. De Lucia,
``Poisoning Attacks and Data Sanitization Mitigations for Machine Learning Models in Network Intrusion Detection Systems,''
in \textit{Proc. 2021 IEEE Military Communications Conference (MILCOM)}, pp.874--879, 2021,
doi: 10.1109/MILCOM52596.2021.9652916.

\bibitem{model_agnostic_mitigation}
G. Severi, S. Boboila, J. Holodnak, K. Kratkiewicz, R. Izmailov, M. J. De Lucia, and A. Oprea,
``Model-Agnostic Clean-Label Backdoor Mitigation in Cybersecurity Environments,''
arXiv preprint arXiv:2407.08159, 2024,
doi: 10.48550/arXiv.2407.08159.

\bibitem{countermeasures}
A. Paracha, J. Arshad, M. B. Farah, and K. Ismail,
``Machine Learning Security and Privacy: A Review of Threats and Countermeasures,''
EURASIP Journal on Information Security, vol.2024, no.10, 2024,
doi: 10.1186/s13635-024-00158-3.

\bibitem{partition}
J. Lee, Y. Cho, R. Lee, S. Yuk, J. Youn, H. Park, and D. Shin,
``A Novel Data Sanitization Method Based on Dynamic Dataset Partition and Inspection Against Data Poisoning Attacks,''
Electronics, vol.14, no.2, article 374, 2025,
doi: 10.3390/electronics14020374.

\bibitem{yudin_retraining}
M. Yudin, A. Reddy, S. Venkatesan, and R. Izmailov,
``Backdoor Attacks During Retraining of Machine Learning Models: A Mitigation Approach,''
in \textit{Proc. 21st International Conference on Security and Cryptography (SECRYPT)}, pp.140--150, 2024,
doi: 10.5220/0012761600003767.

\bibitem{isolation_forest}
F. T. Liu, K. M. Ting, and Z.-H. Zhou,
``Isolation Forest,''
in \textit{Proc. 2008 IEEE International Conference on Data Mining (ICDM)}, pp.413--422, 2008,
doi: 10.1109/ICDM.2008.17.

\bibitem{hdbscan}
R. J. G. B. Campello, D. Moulavi, and J. Sander,
``Density-Based Clustering Based on Hierarchical Density Estimates,''
in \textit{Proc. Pacific-Asia Conference on Knowledge Discovery and Data Mining (PAKDD)}, pp.160--172, 2013,
doi: 10.1007/978-3-642-37456-2\_14.

\bibitem{spectral_signature}
B. Tran, J. Li, and A. Madry,
``Spectral Signatures in Backdoor Attacks,''
in \textit{Proc. 32nd International Conference on Neural Information Processing Systems (NeurIPS)}, pp.8011--8021, 2018.

\bibitem{zheng_clean_label}
W. Zheng and K. Omote,
``Clean-label Backdoor Attack on Machine Learning-based Malware Detection Models and Countermeasures,''
in \textit{Proc. 2022 IEEE International Conference on Trust, Security and Privacy in Computing and Communications (TrustCom)}, pp.1235--1242, 2022,
doi: 10.1109/TrustCom56396.2022.00171.

\bibitem{jigsaw}
L. Yang, Z. Chen, J. Cortellazzi, F. Pendlebury, K. Tu, F. Pierazzi, L. Cavallaro, and G. Wang,
``Jigsaw Puzzle: Selective Backdoor Attack to Subvert Malware Classifiers,''
in \textit{Proc. 2023 IEEE Symposium on Security and Privacy (SP)}, pp.719--736, 2023,
doi: 10.1109/SP46215.2023.10179347.

\end{thebibliography}
\end{document}